\documentclass[9pt,preprint2,natbib209]{aastex61}
\usepackage{graphicx}% Include figure files
\usepackage{dcolumn}% Align table columns on decimal point
\usepackage{bm}% bold math

\def\apjs{ApJS}

\def\apjl{ApJL}

\def\aap{A\&A}

\begin{document}


\title{The Formation of Bimodal Dust Species in Nova Ejecta}

\email{guolianglv@xao.ac.cn}
\author{Adili Duolikun}
\affil{School of Physical Science and Technology,
Xinjiang University, Urumqi, 830046, China}
\author{Chunhua Zhu}
\affil{School of Physical Science and Technology,
Xinjiang University, Urumqi, 830046, China}
\affil{Center for Theoretical Physics,
Xinjiang University, Urumqi, 830046, China}

\author{Zhaojun Wang}
\affil{School of Physical Science and Technology,
Xinjiang University, Urumqi, 830046, China}
\affil{Center for Theoretical Physics,
Xinjiang University, Urumqi, 830046, China}

\author{Helei Liu}
\affil{School of Physical Science and Technology,
Xinjiang University, Urumqi, 830046, China}
\affil{Center for Theoretical Physics,
Xinjiang University, Urumqi, 830046, China}

\author{Lin Li}
\affil{School of Physical Science and Technology,
Xinjiang University, Urumqi, 830046, China}
\affil{Center for Theoretical Physics,
Xinjiang University, Urumqi, 830046, China}

\author{Jinzhong Liu}
\affil{National Astronomical Observatories / Xinjiang Observatory,
the Chinese Academy of Sciences, Urumqi, 830011, China}

\author{Guoliang L\"{u}}
\affil{School of Physical Science and Technology,
Xinjiang University, Urumqi, 830046, China}

%\altaffiltext{3}{Zentrum f\"{u}r Astronomie, Institut f\"{u}r
%Theoretische Astrophysik, Universit\"{a}t Heidelberg,
%Albert-\"{U}berle-Str. 2, D-69120 Heidelberg, Germany.}

\date{\today}

%\pagerange{\pageref{firstpage}--\pageref{lastpage}} \pubyear{2007}

%\maketitle

%\label{firstpage}

\begin{abstract}
The formation of bimodal dust species (namely the silicate and amorphous
carbon dust grains coexistent) in a nova eruption is an open problem.
According to the nova model simulated by Modules for Experiments in Stellar Astrophysics code,
we calculate the formation and growth carbon (C) and forsterite (Mg$_2$SiO$_4$) dust grains
in nova ejecta for the free-expansion model and the
radiative shock model, respectively. In the free-expansion model, the nova ejecta is not an idea
environment for dust nucleation. However, it can efficiently produce
dust in the radiative shock model. We estimate that every nova can produce C grains with
an average mass of about $10^{-9}$ and $10^{-8}$ ${\rm M_\odot}$,
and Mg$_2$SiO$_4$ grains with an average mass of about $10^{-8}$ and $10^{-7}$ ${\rm M_\odot}$.
Based on the mass of ejected gas, the ratio of dust to gas is about 1\%.
The C grains form first after several or tens of days of nova eruption.
After that, the Mg$_2$SiO$_4$ grains begin to grow in tens of days,
which is consistent with observations.
\end{abstract}

\section{Introduction}
As it is well known, the interstellar medium (ISM) is made
up of gas and dust.  The latter offers an unique probe of the ISM across multiple size,
density, and temperature scales.
Based on the popular view of point, dust is produced in the stellar outflows, such as
the stellar winds from asymptotic giant branch (AGB) star, red supergiant star, Wolf-Raynet star and OB star,
the ejecta from planetary nebula, supernova (SN), nova and common envelope
\citep[e. g.,][]{Ferrarotti2006,Ventura2014,Todini2001,Barlow2010,Zhukovska2016,Lu2013,Zhu2013,Zhu2015}.
Due to the many poorly understood processes evolved in dust formation and growth, our knowledge of it
is extremely limited \cite[e. g.,][]{Gail1999,Draine2009}.
Generally, the stellar wind of AGB star and the ejecta of SN are considered the main dust sources,
while others only offer a small fraction of interstellar dust\citep{Tielens2005,Zhu2015,Zhukovska2016}.
Especially, the contribution of the novae to the Galactic dust is insignificant\citep{Draine2009}.

However, nova is an excellent laboratory for investigating dust formation.
\cite{Harrison2018} considered that about 50\% of nova eruptions can produce
dust.
Compared with SN, nova eruption has higher occurrent rate ($\sim 100$ yr$^{-1}$)
and closer distance in the Galaxy\citep{Lu2009,Draine2009,Li2016PASJ,Rukeya2017}.
Compared with AGB star whose thick stellar wind obscures the dust formation,
many novae can produce dust during every eruption.
Especially, some novae (V1370 Aql, V842 Cen, QV Vul, V2676 Oph,
V1280 Sco and V1065 Cent) can successively produce amorphous carbon and silicate dust
grains during an eruption\citep{Strope2010,Helton2010,Sakon2016,Kawakita2017}.

The formation of bimodal dust species (that is, morphous carbon and silicate dust
grains) is still debated\citep{Sakon2016}. Very recently, \cite{Zhu2019} simulated
the evolution of the abundance ratio of the carbon to the oxygen ($C/O$) in nova ejecta during an whole eruption,
and found that some nova ejecta is an ideal chemical environment for the formation of bimodal dust species.
However, they do not discuss the physical conditions for it.
Usually, the environment of a nova eject is not an ideal environment for forming dust.
Recently, \cite{Derdzinski2017} suggested that the radiative shocks in nova ejecta can offer
the environments for dust formation.

In this paper, combining the shock model of \cite{Derdzinski2017} and the nova model of \cite{Zhu2019},
we investigate the formation of bimodal dust species in nova ejecta.
In \S 2, we present our assumptions and describe some details of the modelling algorithm. In Section 3, we discuss the main
results and the effects of different parameters. In Section 4, the main
conclusions are given.

\section{Model}
In order to simulate the dust formation in the nova ejecta,
we must construct a model including nova eruption, the ejecta expansion and
dust nucleation.

\subsection{Nova}
Since \cite{Starrfield1972} first simulate the thermal nuclear runaway (TNR) of a nova by a nuclear reaction network,
there are many theoretical models for nova eruption\citep[e. g.][]{Prialnik1995,Jose1998,Yaron2005,Glasner2012,Casanova2016,Casanova2018}.
Modules for Experiments in Stellar Evolution (MESA, [rev. 10108]; \cite{Paxton2011,Paxton2013,Paxton2015,Paxton2018}) also offers
a model for calculating nova eruption,  which has been used in \cite[e. g.,][]{Denissenkov2013,Denissenkov2014}.
\cite{Zhu2019} used MESA to investigate the evolution of the chemical compositions in nova ejecta. They found that nova ejecta
may offer a chemical environment for the formation of bimodal dust species.

The present paper use the nova model of \cite{Zhu2019}, in which the nova eruptions are affected by input parameters
as below: WD mass, mass-accretion rate and mixing depth ($\delta=\frac{M_{\rm mix}}{M_{\rm WD}}$, where $M_{\rm mix}$ is the
mixed mass of WD and $M_{\rm WD}$ is the WD mass).
Here, following \cite{Zhu2019}, we discuss the effects of these input parameters on the formation of bimodal dust species.
However, as shown in many literatures\citep[e. g.][]{Glasner2012,Casanova2016,Zhu2019}, the mixing between the accreted matter and
the underlying WD material only occurs a very thin envelope close to WD surface. Therefore, we take a $\delta$ with a small value of
0.001 in this work.

In addition, nova eruption is also affected by the core temperatures of WDs\citep[e. g.][]{Jose1998,Yaron2005}.
The cooling model of WD depends on the atmospheric treatment, the convection, the radiative transfer,
crystallization, and so on\citep[e. g.][]{Wood1992,Hansen1999,Liu2019}.
In the present work, we do not the effects of WD core temperature.

\subsection{Ejecta Expansion}
When nova eruption occurs, the TNR ash is blown away from WD.
The matter ejected begins to expand. The evolution of density and
temperature of the ejecta is crucial for dust formation.
They depend on the model of nova expansion.

If the nova ejecta
freely expands and is ideal gas, the time evolution of gas density and temperature in nova ejecta is
similar with the model in \cite{Nozawa2003}.
The evolution of gas density is given by
\begin{equation}
\rho(t)=\rho_0(t) (\frac{t}{t_0})^{-3},
\label{eq:rho}
\end{equation}
where $t_0$ is 1 day after nova eruption, and $\rho_0$ is the initial density.
The temperature evolution is given by
\begin{equation}
T(t)=T_0(t) (\frac{t}{t_0})^{3(1-\gamma)}.
\label{eq:tem}
\end{equation}
Following \cite{Fransson1989} and \cite{Kozasa1989}, the parameter
$\gamma$ in this work is taken as 1.25.

If there is no interaction (such as wind collision) in the ejecta,
nova mainly offers the eruption energy in the optical spectra produced by TNR.
However, the emissions in the high-energy spectra are observed during some nova outbursts.
\cite{Mukai2008} suggested that all novae are transient hard X-ray sources
powered by shocks within the ejected shell.
Using the Fermi Large Area Telescope,
\cite{Abdo2010} reported that the nova of SS V407 Cygni had variable $\gamma$-ray emission (0.1-10 GeV).
Up to now, there are nine novae with $\gamma$-ray emissions during their outbursts\citep{Ackermann2014,Cheung2016,Li2016}.
The shock model can explain the nova $\gamma$-ray emission detected by Fermi Large Area Telescope\citep{Abdo2010,Lu2011,Martin2013,Sun2016,Martin2018}.
Therefore, shocks play an important role in the nova eruption.
\cite{Metzger2014} investigated the shocks triggered by the interaction between the fast nova outflow and a dense circumstellar shell,
and found that these shocks may be radiative when the density of nova ejecta becomes very high\citep[Also see][]{Metzger2015}.
This radiation makes the post-shock gas efficiently cooled and simultaneously enhance its density by a factor of $\leq 10^3$\citep{Metzger2014,Metzger2015}.
A cool and dense shell is produced between the forward and reverse shocks.
Following \cite{Metzger2014} and \cite{Metzger2014}, \cite{Derdzinski2017} gave the
characteristic density of the cold shell by
\begin{equation}
n_{\rm max}\approx4\times10^{14}t^{-3}_{\rm wk}v^{-1}_8M_{-4}T_{\rm CS, 4}\ cm^{-3},
\label{eq:nmax}
\end{equation}
where, $t_{\rm wk}$ is the time in weeks, $v_8=v_{\rm sh}/(10^8 {\rm cm s^{-1}})$, $M_{-4}=M_{\rm ej}/(10^{-4} M_\odot)$ and
$T_{\rm CS, 4}=T_{\rm CS}/(10^4 {\rm K})$, respectively. Here, $v_{\rm sh}$ is the shock velocity, $M_{\rm ej}$ is
the ejecta mass and $T_{\rm CS}$ is the temperature of cold shell.
Considering the radiative heating\citep{Pontefract2004}, they gave the the temperature of cold shell by
\begin{equation}
T_{\rm CS}\approx2500{\rm K}\frac{L}{10^{38}{\rm erg s^{-1}}}v^{-1/2}_8t^{-1/2}_{\rm wk}.
\label{eq:tcs}
\end{equation}
Using the above shock model, \cite{Derdzinski2017} studied the dust formation in this shell.
They found that the dust grains can grow efficiently to
large sizes ($\leq 0.1\mu$m), which is consistent with the observations\citep{Gehrz1998,Sakon2016}.
However, they only considered the dust nucleation in a nova ejecta with a fixed chemical compositions although
they also changed $C/O$.
As shown in  \cite{Denissenkov2014} and \cite{Zhu2019}, the chemical compositions of the ejecta during a whole
nova eruption is ever-changing, which can result in $C/O$ with varied values.

In this work, we use the free expansion model and shock model to investigate
the formation of bimodal dust species in the nova ejecta, respectively.
In fact, there is the pre-existing circum stellar medium when the nova ejecta expands.
Its chemical properties also affect the element abundances and $C/O$ of the cool and dense shell
between the forward and reverse shocks. However, it is very difficult to determine the
chemical properties of the pre-existing circum stellar medium because it may originate
from the ejecta of nova eruption, or the matter transferred from WD companion, or the mixing of them.
As far as we know, there is no any observational evidence or theoretical model referring to it.
Therefore, in this work, following \cite{Derdzinski2017}, we only consider a shock model within nova ejecta.
In addition, nova expansion model also depends on the geometry of nova wind and the pre-existing cirumstellar medium,
which are usually not spherically symmetric. For example, \cite{Chomiuk2014} found that,
due to the motion of binary system in nova V959Mon, the denser material was expelled out along
the equatorial plane while the more thin gas was ejected rapidly along the poles from WD.
However, for simplicity, we assume a spherically symmetric ejecta in this work.

\subsection{Dust Nucleation}
As the last section discusses, it is possible that
the bimodal dust species are produced in a nova eruption.
However, based on the classical nucleation theory,
the dust grains can not form until a gas is supersaturated \citep{Becker1935,Feder1966}.
Following \cite{Derdzinski2017}, we only consider the possibility for
the formation of carbon grains and Mg$_2$SiO$_4$ which represents the silicate grain
population during a nova eruption.

In the classical nucleation theory, it is determined by the ratio of the gas density
to the equilibrium density ($n_{\rm eq}$) whether a gas becomes supersaturated.
For carbon grains, this ratio is given by
\begin{equation}
S_{\rm C}=n_{\rm C}/n_{\rm eq},
\label{eq:sc}
\end{equation}
where $n_{\rm C}$ is the number density of carbon in the gas state, and
$n_{\rm eq}=\frac{6.9\times10^{13}}{k_{\rm B}T}e^{-84428.2/T}$ \citep{Keith2011}.
Here, $k_{\rm B}$ and $T$ are the Boltzmann constant and the gas temperature,
respectively.
For forsterite grains, whose formation involves several elements (2Mg + SiO + 3O $\rightarrow$ Mg$_2$SiO$_4$),
the ratio is given by,
\begin{equation}
\ln S_{\rm Si}=-\frac{\Delta G}{K_{\rm B}T}+\sum_{\rm i}v_{\rm i}\ln p_{\rm i},
\label{eq:sic}
\end{equation}
where $\Delta G$ and $v_{\rm i}$ are Gibbs free energy for the chemical reaction and
the stoichiometric coefficients, respectively. Their values can be found in \cite{Kozasa1987} and \cite{Nozawa2003}.
The $p_{\rm i}$ are the partial pressures of each species.

When $S_{\rm Si}$ or $S_{\rm C}$ is greater than 1, the dust nucleation occurs. There are many
models for dust nucleation and growth\citep[e. g.,][]{Gail1999,Ferrarotti2006,Ventura2012,Todini2001,Bianchi2007}.
For simplicity, following \cite{Derdzinski2017}, we only discuss the formation and growth of two dust species --- solid
C and Mg$_2$SiO$_4$. We assumed that a density of seed nuclei per hydrogen nucleus, $n_{\rm d}$, is $10^{-13}$ and
the the radius of the seed nuclei, $a_0$, is $10^{-7}$ cm \citep{Gail1999}.
The dust grains continuously grow by accreting at a rate\citep{Derdzinski2017}
\begin{equation}
\frac{{\rm d}a}{{\rm d}t}=n^{\rm gas}c_{\rm cs}(a_0v)^{2/3}\sqrt{(\frac{k_{\rm B}T}{2\pi m})}\ {\rm s}^{-1},
\label{eq:grow}
\end{equation}
where $n^{\rm gas}$ is the number density of C or Si atoms in gas phase, $v_{\rm i}$ is the volume of
solid C or Mg$_2$SiO$_4$, $m$ is the mass of the molecule and $c_{\rm cs}=(30\pi)^{1/3}$ for spherical grains.
Of course, when the grains grow,  evaporation and chemisputtering can occur. In this work, we
neglect them.

\section{Results}
\cite{Zhu2019} had taken into account 48 different model combinations (4 WD masses, 4 mixing depths, 3 WD mass-accretion rates).
In their work, a mixing depth $\delta$ larger than 0.05 can result in a very low $C/O$ because the O abundance of
the WD from the surface to the inside quickly rises above C abundances (See Figure 1 in \cite{Zhu2019}).
In these models, $C/O$ is always lower than 1. For the models with $\delta=$ 0.001, $C/O$ of nova ejecta over the course of an eruption
can evolve from greater than 1.0 to less than 1.0. That is, the models with a small mixing depth of 0.001
can offer the chemical conditions for the formation of bimodal dust species which is the focal point of this work.
Therefore, we take $\delta$ as 0.001.
 We choose the two typical models as below: novae for 1.0 $M_\odot$ CO WD and
1.2 $M_\odot$ ONe WD with the mixing depth of 0.001 and different mass-accretion rates of
$10^{-7}$ and $10^{-9} M_\odot$ yr$^{-1}$, respectively.

\begin{figure}
\includegraphics[totalheight=3.5in,width=3.0in,angle=-90]{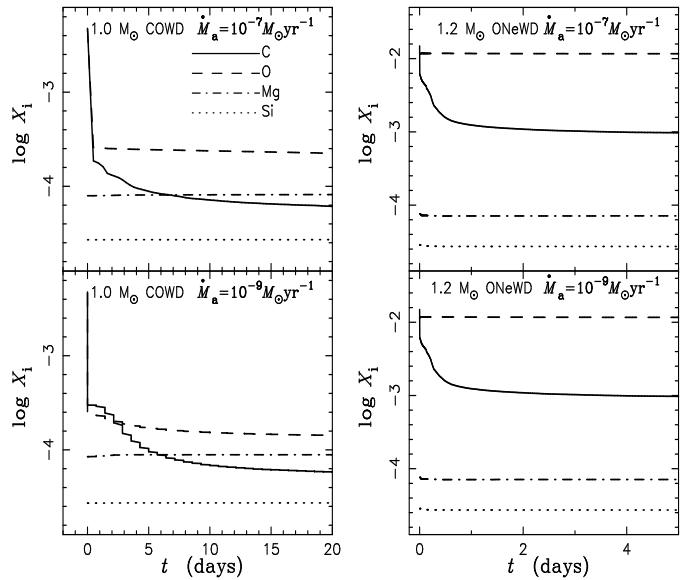}
\caption{Evolution of normalized number density for C, O, Mg and Si elements during an entire eruption.
The input parameters of models are given in the top of every panel.
The solid, dashed and dash-doted and doted lines represent the C, O, Mg and Si elements, respectively.}
\label{fig:cot}
\end{figure}

Figure \ref{fig:cot} shows the evolution of normalized number density for C, O, Mg and Si
elements during an entire eruption.
Because C abundance around the surface of WD is higher than O abundance, C/O of ejecta at the beginning of nova eruption is larger than 1.
However, the TNR rapidly depletes C element, and results in the C/O less than 1.
As discussed in \cite{Zhu2019}, it is very possible for bimodal dust species to form in such ejecta.

\begin{figure}
\includegraphics[totalheight=3.5in,width=3.0in,angle=-90]{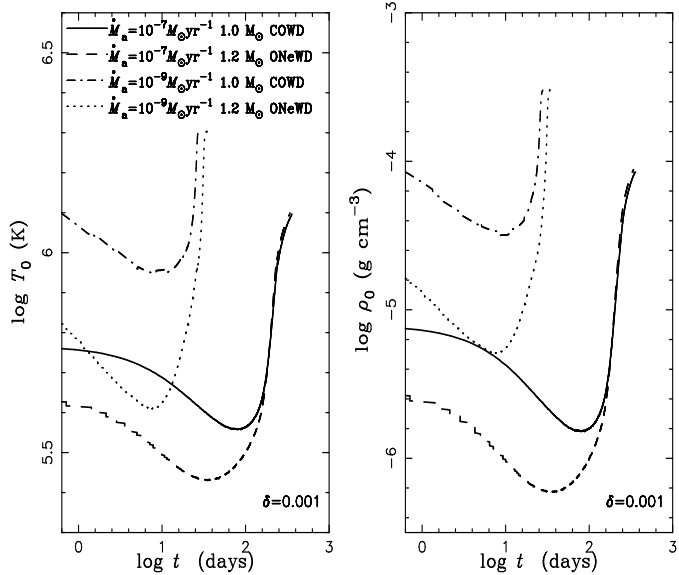}
\caption{The initial temperature and density of gas ejected by TNR.
The different line styles representing the different models are showed the top of
left panel.}
\label{fig:trou0}
\end{figure}

Figure \ref{fig:trou0} shows the initial temperature and density of gas ejected by TNR.
Not only input parameters but also the ejected time affect these physical quantities
which directly determine the dust formation.
Therefore, we choose the ejected gas
at the beginning of nova eruption, at the maximum luminosity and at
the end of nova eruption, to investigate the possibility of dust nucleation by
using the models of ejecta expansion.

\begin{figure}
\includegraphics[totalheight=3.5in,width=3.0in,angle=-90]{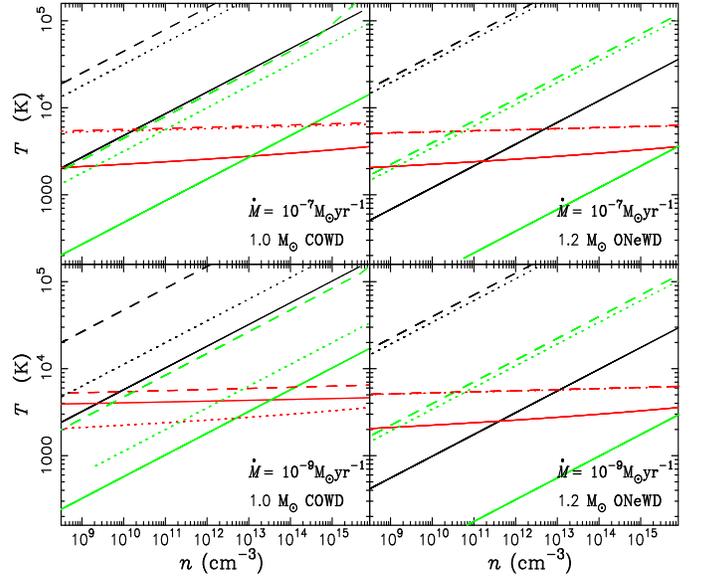}
\caption{Number density and temperature evolution of nova ejecta.
Every panel represents a model whose parameters are given in the
right-bottom region and the mixing depth equals 0.001.
The solid, dashed and dot-dashed lines give the results for the ejected gas
at the beginning of nova eruption, at the maximum luminosity and at
the end of nova eruption, respectively. The black and green lines represent
the models of free expansion and radiative shock, respectively. The red lines
show the relation of temperature and its equilibrium density ($n_{\rm eq}$). }
\label{fig:rouev}
\end{figure}

Figure \ref{fig:rouev} gives the evolution
of gas number density in  the models of free expansion and radiative shock
and its equilibrium density with the temperature.
The region in which $n_{\rm C}>n_{\rm eq}$ or $n_{\rm Si}>n_{\rm eq}$ is
favorable to dust nucleation. For the model of free expansion, we find that
dust is hardly produced in these regions for all phases,  no matter at the beginning of nova eruption,
at the maximum luminosity or at the end of nova eruption.
There are too high temperature or too low number density in these regions.
For example, for the model with 1.0 $M_\odot$ CO WD and a mass-accretion rate of $10^{-7} M_\odot$yr$^{-1}$,
that is, the model is showed by the solid black and red lines in the left-top panel of Figure \ref{fig:rouev},
the temperature of the gas is about 2000 K when the number density is higher than the equilibrium density,
and simultaneously the number density of free carbon is only about $2.5\times10^8$ cm$^{-3}$.
In such environment, the nucleation and growth of dust grains hardly occur \citep[e.g.,][]{Gail1986}.
Therefore, the free expansion model is unsuitable for dust formation in nova ejecta.

For the model of radiative shock, the density under shock compression
can be enhanced by about 3$-$5 magnitude \citep{Derdzinski2017}.
Therefore, as Figure \ref{fig:rouev} shows, when  $n_{\rm C}>n_{\rm eq}$ or $n_{\rm Si}>n_{\rm eq}$,
the temperature and density of ejected gas in radiative shock are about 2000 K and $10^{12}$---$10^{13}$ cm$^{-3}$.
The dust nucleation and growth may occur in this environment.
Therefore, we only consider the dust formation in the model of radiative shock.

\begin{figure}
\includegraphics[totalheight=3.5in,width=3.0in,angle=-90]{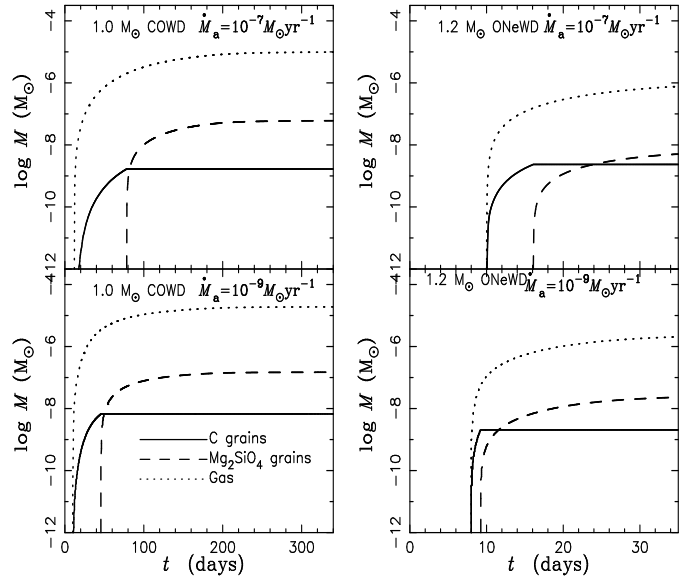}
\caption{The dust yields in nova ejecta. The solid and dashed lines represent
the yields of C and Mg$_2$SiO$_4$ grains, respectively. The dotted lines give
the mass of ejected gas. The input parameters of models are given in the top of
every panel.}
\label{fig:gdust}
\end{figure}

Figure \ref{fig:gdust} shows the dust yields calculated by Eqs. (\ref{eq:nmax}), (\ref{eq:tcs}) and (\ref{eq:grow}).
Because the $C/O$ of ejecta is higher than 1 at the beginning of the nova eruption (See Figure \ref{fig:cot}),
the C grains form first. After several or tens of days, the Mg$_2$SiO$_4$ grains begin to grow.
On the observations, the dust formation occurs in about 20---100 days after nova eruption\citep[e. g.,][]{Geisel1970,Gehrz1980,Evans2012,Raj2016}.
Our results are consistent with observations.
In our models, every nova can produce C grains with an average mass of about $10^{-9}$ and $10^{-8}$ ${\rm M_\odot}$,
and Mg$_2$SiO$_4$ grains with an average mass of about $10^{-8}$ and $10^{-7}$ ${\rm M_\odot}$.
Based on the mass of ejected gas, the ratio of dust to gas is about 1\%.
From the observational point of view, majority of the dusty classical novae are COWD novae
and only a few exceptional ONeMg novae have shown the signs of dust formation,
e.g., V1370 Aql, V838 Her and V1065 Cen \citep{Gehrz1984,Woodward1992,Helton2010}.
The main reason, showed by Figure \ref{fig:gdust}, is that ONeWD usually has more mass than COWD.
Under similar input parameters, the critical mass accreted by ONeWD for nova eruption is smaller than that by COWD.
Then, the mass ejected by the former is lower than that by the later.
In our simulations, the mass ejected by COWD nova is about 10 times higher than that by ONeWD nova.
In turn, COWD nova can produce more dust grains than ONeWD. Therefore, the possibility of observing dust grains
in COWD nova is higher than ONeWD nova, which is consistent with observations.

\section{Conclusions}
Using the nova model provided by MESA, we investigate
the possibility of nova ejecta producing bimodal dust species.
We find that it is very hardly difficult to form dust when nova ejecta freely expands.
However, if the radiative shock occurs, the nova ejecta can efficiently
produce dust. In our models, every nova can produce C grains with an
average mass of about $10^{-9}$ and $10^{-8}$ ${\rm M_\odot}$,
and Mg$_2$SiO$_4$ grains with an average mass of about $10^{-8}$ and $10^{-7}$ ${\rm M_\odot}$.
Based on the mass of ejected gas, the ratio of dust to gas is about 1\%.

Obviously, in the radiative shock model, nova ejecta offers a suitable environment for dust nucleation.
However, as discussed in \cite{Gail1999}, our knowledge to dust formation and growth is still extremely limited.
Especially, in this work, we only consider C and  Mg$_2$SiO$_4$ grains. In fact,
based on the chemical environment, nova ejecta can produce many species of dust, such as
olivine-type, pyroxene-type, quartz-type, iron, SiC-type dust grains, and so on\citep[e. g.,][]{Ferrarotti2006}.
There is still long way to understand the formation and growth of bimodal species dust in the nova ejecta.

\section*{Acknowledgments}
This work received the generous support of the  National Natural Science Foundation of China,
project Nos. 11763007, 11473024, 11463005, 11863005, 11803026
and 11503008. We would also like to express our gratitude to the Tianshan Youth Project of Xinjiang No.2017Q014.

%\newpage %Just because of unusual number of tables stacked at end
\bibliography{lglmn,lglapj}
%\begin{thebibliography}{99}
%\include{lgl.bib}
%\end{thebibliography}

\label{lastpage}

\end{document}